\documentclass[useAMS,referee,usenatbib]{biom}

\usepackage{amssymb}
\usepackage{amsmath}
\usepackage{algorithm}
\usepackage{algpseudocode}
\usepackage{graphicx}

\def\bSig\mathbf{\Sigma}

\title[This is an Example of Recto Running Head]{Biomarkers selection and combination based on the weighted Youden index}

\author
{Ao Sun\emailx{quater\_sa@stu.pku.edu.cn} \\
Academy for Advanced Interdisciplinary Studies, Peking University, Beijing, China
\and
Zhanwang Deng\emailx{dzw\_opt2022@stu.pku.edu.cn} \\
Academy for Advanced Interdisciplinary Studies, Peking University, Beijing, China
\and
Jiahui Zhao\emailx{zzjh\_2007@163.com} \\
Department of Dermatology, Peking University First Hospital, Beijing, China
\and
Hang Li\emailx{drlihang@126.com} \\
Department of Dermatology, Peking University First Hospital, Beijing, China
\and
Xiao-Hua Zhou\emailx{azhou@math.pku.edu.cn} \\
Beijing International Center for Mathematical Research and Department of Biostatistics, \\
Peking University, Beijing, China
}

\begin{document}

\pagerange{\pageref{firstpage}--\pageref{lastpage}} 
\volume{64}
\pubyear{2008}
\artmonth{December}

\label{firstpage}


\begin{abstract}
In clinical practice, multiple biomarkers are used for disease diagnosis, but their individual accuracies are often suboptimal, with only a few proving directly relevant. Effectively selecting and combining biomarkers can significantly improve diagnostic accuracy. Existing methods often optimize metrics like the Area Under the ROC Curve (AUC) or the Youden index. However, optimizing AUC does not yield estimates for optimal cutoff values, and the Youden index assumes equal weighting of sensitivity and specificity, which may not reflect clinical priorities where these metrics are weighted differently. This highlights the need for methods that can flexibly accommodate such requirements. In this paper, we present a novel framework for selecting and combining biomarkers to maximize a weighted version of the Youden index. We introduce a smoothed estimator based on the weighted Youden index and propose a penalized version using the SCAD penalty to enhance variable selection. To handle the non-convexity of the objective function and the non-smoothness of the penalty, we develop an efficient algorithm, also applicable to other non-convex optimization problems. Simulation studies demonstrate the performance and efficiency of our method, and we apply it to construct a diagnostic scale for dermatitis.
\end{abstract}

%



\maketitle

\section{Introduction}
\label{sec:intro}

Accurate disease prediction and classification based on clinical and laboratory data are crucial for improving diagnostic precision in medical research and practice. While individual diagnostic tests or biomarkers often lack sufficient reliability, integrating multiple biomarkers has emerged as a common strategy to enhance diagnostic performance. Linear combinations of biomarkers are particularly attractive due to their simplicity, interpretability, and robust performance \citep{pepe2006combining}. However, challenges arise when large numbers of biomarkers are measured, yet only a small subset is typically relevant to clinical outcomes. This underscores the importance of effective biomarker selection methods, which reduce data dimensionality, improve interpretability, and facilitate the development of compact diagnostic tools.

Existing approaches for biomarker selection and combination frequently utilize logistic regression or support vector machines (SVM) \citep{li2022applying, becker2011elastic}, valued for their simplicity and robustness. However, these methods are limited by their inability to directly optimize receiver operating characteristic (ROC)-related metrics, which are pivotal in medical diagnostics. To address this gap, increasing attention has been directed toward methods that explicitly target ROC-based objectives, such as the Area Under the Curve (AUC), partial AUC (pAUC), and the Youden index—key metrics for assessing the performance of diagnostic models.

Approaches targeting ROC-related metrics can be broadly categorized into two main types. The first category comprises empirical performance-based methods, which typically follow a two-step framework: biomarkers are first selected based on their contributions to metrics like AUC or pAUC, and then combined. For example, \citet{sun2017avc} proposed maximizing AUC through feature complementarity, while \citet{hsu2014biomarker} optimized pAUC using stepwise selection. Although practical and easy to implement, these methods often lack rigorous theoretical foundations \citep{breiman1996heuristics} and are prone to overfitting in high-dimensional settings, resulting in unstable results.

The second category involves regularization-based approaches that integrate biomarker selection and model fitting into a unified framework. By assigning zero coefficients to irrelevant features, these methods achieve simultaneous feature selection and optimization, reducing overfitting and enhancing computational efficiency. Notable examples include the TGDR algorithm \citep{ma2005regularized}, penalized regression models \citep{yu2014aucpr}, and SCAD penalization for AUC or C-statistics \citep{ma2017concordance, lin2011selection}. However, these methods often rely on restrictive assumptions, such as multivariate normality or single-index models, which may not hold in real-world applications. Additionally, optimization algorithms commonly employed in these methods, such as local quadratic approximation \citep{fan2001variable} and Newton-Raphson methods, face significant challenges in non-convex settings. Their dependence on carefully chosen initial points further limits robustness.

More recently, \citet{salaroli2023pye} introduced a penalized Youden index-based method that incorporates various penalty terms. While innovative, this approach lacks theoretical guarantees and employs a generic optimization algorithm (MAPG) that is not specifically tailored for biomarker selection tasks, reducing its practical effectiveness.

Despite their utility, metrics like AUC and pAUC lack granularity for clinical decision-making because they do not provide explicit cutoff points. The Youden index, by identifying an optimal cutoff, addresses this limitation but assumes equal importance for sensitivity and specificity, which may not align with clinical priorities. For instance, high sensitivity is critical for diagnosing severe conditions like pheochromocytoma, where missed diagnoses could lead to severe consequences \citep{unger2006diagnostic}.

To address these limitations, we introduce a novel approach that combines practical applicability, theoretical rigor, and computational innovation. By adopting the weighted Youden index as the optimization objective, our method allows for greater flexibility in aligning with clinical priorities, such as balancing sensitivity and specificity. Unlike existing methods that rely on restrictive assumptions, such as single-index models or multivariate normality, our approach employs a more general framework ensuring parameter uniqueness. Theoretically, we establish the consistency and oracle properties of the smoothed weighted Youden index estimators and the proposed penalized estimator, effectively addressing key approximation challenges and parameter restrictions in the proof. To optimize the proposed model, we develop a novel nonmonotone accelerated proximal gradient (NAPG) algorithm tailored for this non-convex optimization. This algorithm integrates polynomial interpolation-based backtracking line search, proximal gradient updates, nonmonotonicity, and dynamic step-size control to achieve faster convergence and improved robustness. Beyond our primary application, the NAPG algorithm is versatile and can be applied to other non-convex problems, such as optimizing AUC-based objectives, further enhancing its utility.

The remainder of this paper is organized as follows. Section 2 introduces the notation and theoretical foundations of the smoothed weighted Youden index. Section 3 presents the penalized estimator and its theoretical properties. Section 4 describes the optimization algorithm, while Section 5 reports simulation results. Section 6 illustrates a real-world application, and Section 7 concludes with a discussion of future extensions.

\section{Preliminaries}
\subsection{Notations}
Consider a diagnostic scenario involving $n$ patients, divided into two groups: diseased $(D=1)$ and healthy $(D=0)$. Let $n_1$ and $n_0$ denote the number of diseased and healthy individuals, respectively, with $n=n_0+n_1$ representing the total sample size. Each patient is assessed using $p$ continuous biomarkers, collectively represented as $\mathbf{T}$. For the diseased group, $\mathbf{X}$ denotes the $p$-dimensional vector of biomarker measurements, where $\mathbf{X}_i=(X_{i1},\cdots,X_{ip})^T$ corresponds to the biomarker values for the $i$th diseased subject $(i=1,\cdots,n_1)$. Similarly, for the healthy group, $\mathbf{Y}$ represents the $p-$dimensional biomarker vector, where $\mathbf{Y}_j=(Y_{j1},\cdots,Y_{jp})^T$ corresponds to the biomarker values for the $j$th healthy subject $(j=1,\cdots,n_0)$.

To enhance diagnostic accuracy, multiple biomarkers are often linearly combined. Given a set of linear combination coefficients $\mathbf{\omega}$ and a cutoff point $c$, the sensitivity of the linear predictor $\mathbf{\omega}^T\mathbf{T}$ is defined as $Se_{\mathbf{\omega}}(c)=\mathbb{P}\left(\mathbf{\omega}^T\mathbf{T}> c\mid D=1\right)=\mathbb{P}\left(\mathbf{\omega}^T\mathbf{X}> c\right)$, and its specificity is defined as $Sp_{\mathbf{\omega}}(c)=\mathbb{P}\left(\mathbf{\omega}^T\mathbf{T}\leq c\mid D=0\right)=\mathbb{P}\left(\mathbf{\omega}^T\mathbf{Y}\leq c\right)$. For a pre-specified weight $\pi\in(0,1)$, the weighted Youden index is given by \citet{li2013weighted}:
\begin{equation}
    \begin{split}
        J_{\pi,\mathbf{\omega}}&=\max_{c}\left\{2\left(\pi Se_{\mathbf{\omega}}(c)+(1-\pi)Sp_{\mathbf{\omega}}(c)\right)-1\right\}.
    \end{split}
\end{equation}
When $\pi=0.5$, this reduces to the classical Youden index:
$$J_{\mathbf{\omega}}=\max_{c}\left\{Se_{\mathbf{\omega}}(c)+Sp_{\mathbf{\omega}}(c)-1\right\}.$$

This study aims to identify a linear combination of biomarkers and a threshold that maximize the weighted Youden index for a given $\pi$. Mathematically, this can be formulated as:
\begin{equation}
\begin{split}
    \mathbf{\omega}_0,c_0
    &=\arg\max_{\mathbf{\omega}\in\Omega,c\in\mathbb{R}}\left\{2\left((1-\pi)\mathbb{P}(\mathbf{\omega}^T\mathbf{Y}\leq c)-\pi \mathbb{P}(\mathbf{\omega}^T\mathbf{X}\leq c)\right)+2\pi\right\} \\
    &=\arg\max_{\mathbf{\omega}\in\Omega,c\in\mathbb{R}}\left\{(1-\pi)\mathbb{P}(\mathbf{\omega}^T\mathbf{Y}\leq c)-\pi \mathbb{P}(\mathbf{\omega}^T\mathbf{X}\leq c)\right\},
\end{split}
\end{equation}
where $\Omega=\{\mathbf{\omega}\in\mathbb{R}^P:\left\|\mathbf{\omega}\right\|=1\}$ ensures identifiability. The constraint $\left\|\mathbf{\omega}\right\|=1$ avoids redundancy, as the decision rule based on $\alpha_0+\alpha_1\mathbf{\omega}^{\prime}\mathbf{T}>\alpha_0+\alpha_1 c$ is equivalent to $\mathbf{\omega}^{\prime}\mathbf{T}>c$.

\subsection{Estimation}
A nonparametric estimate of $(\mathbf{\omega}_0,c_0)$ is defined as:
\begin{equation} \label{prob-est}
    (\widehat{\mathbf{\omega}}^{\text{E}},\widehat{c}^{\text{E}})=\arg\max_{\mathbf{\omega}\in\Omega,c\in\mathbb{R}}\left\{(1-\pi)\frac{1}{n_0}\sum_{j=1}^{n_0}\mathbb{I}\left(\mathbf{\omega}^{T}\mathbf{Y}_j\leq c\right)-\pi\frac{1}{n_1}\sum_{i=1}^{n_1}\mathbb{I}\left(\mathbf{\omega}^{T}\mathbf{X}_i\leq c\right)\right\},
\end{equation}
here $\mathbb{I}(\cdot)$ is the indicator function. However, the discontinuity introduced by $\mathbb{I}(\cdot)$ makes the objective function non-differentiable, increasing computational complexity, especially in high-dimensional settings where the cost grows exponentially with $n^p$. To simplify computation, we approximate the indicator function with the smooth and differentiable cumulative distribution function of the standard normal distribution, $\Phi(\cdot)$, which is more stable and accurate than alternatives such as the sigmoid function \citep{lin2011selection}. The indicator function $\mathbb{I}\left(\mathbf{\omega}^{T}\mathbf{X}_i\leq c\right)$ is replaced with $\Phi\left(\left(c-\mathbf{\omega}^T\mathbf{X}_i\right)/h_n\right)$, and similarly, $\mathbb{I}\left(\mathbf{\omega}^{T}\mathbf{Y}_j\leq c\right)$ is approximated by $\Phi\left(\left(c-\mathbf{\omega}^T\mathbf{Y}_j\right)/h_n\right)$. Here, $h_n$ is a bandwidth parameter selected to converge to zero. The resulting smoothed weighted Youden index estimator is given by:
\begin{equation}
    (\widehat{\omega},\widehat{c})=\arg\max_{\omega\in\Omega,c\in\mathbb{R}}\left\{(1-\pi)\frac{1}{n_0}\sum_{j=1}^{n_0}\Phi\left(\frac{c-\mathbf{\omega}^{T}\mathbf{Y}_j}{h}\right)-\pi\frac{1}{n_1}\sum_{i=1}^{n_1}\Phi\left(\frac{c-\mathbf{\omega}^{T}\mathbf{X}_i}{h}\right)\right\}.
\end{equation}
This smoothing approach facilitates efficient optimization while maintaining desirable theoretical properties such as consistency and asymptotic normality.

\subsection{Properties of the estimator}

Previous studies (e.g., \citet{pepe2006combining, ma2007combining}) have commonly adopted the single-index model to study estimator properties in biomarker combination problems, assuming that $\mathbb{P}(D = 1 \mid \mathbf{T}) = G(\mathbf{\omega}_0^T \mathbf{T})$, where $G(\cdot)$ is an increasing link function. Under this model, the combination score $\mathbf{\omega}_0^T \mathbf{T}$ yields the optimal receiver operating characteristic (ROC) curve \citep{neyman1933ix}. Specifically, for any fixed false positive rate (FPR), the decision rule $\mathbf{\omega}_0^T \mathbf{T} > c$ achieves a higher true positive rate (TPR) than any other scoring rule with the same FPR. This implies that $\mathbf{\omega}_0^T \mathbf{T}$ maximizes the vertical distance between the ROC curve and the line $y = \frac{1 - \pi}{\pi}x$, which corresponds to maximizing the weighted Youden index. To ensure the uniqueness of the optimal cutoff $c_0$, additional conditions are typically required. A sufficient condition is the existence of a threshold $c_0$ such that $\pi f_{\mathbf{\omega}_0^{T}\mathbf{X}}(c_0)=(1-\pi)f_{\mathbf{\omega}_0^{T}\mathbf{Y}}(c_0)$, with $(1-\pi)f_{\mathbf{\omega}_0^{T}\mathbf{Y}}(t)>\pi f_{\mathbf{\omega}_0^{T}\mathbf{X}}(t)$ for $t<c_0$ and $(1-\pi)f_{\mathbf{\omega}_0^{T}\mathbf{Y}}(t)<\pi f_{\mathbf{\omega}_0^{T}\mathbf{X}}(t)$ for $t>c_0$. These conditions guarantee that $c_0$ uniquely maximizes the weighted Youden index.

However, the single-index model imposes restrictive assumptions. To address this limitation, we propose a more general framework that relaxes the single-index assumption while still ensuring the identifiability of $(\mathbf{\omega}_0, c_0)$ under weaker conditions.

We first establish the existence of a global maximizer of the weighted Youden index:
\begin{theorem}\label{theorem0}[Existence]
    Let $\Omega = \{\mathbf{\omega} \in \mathbb{R}^P : \|\mathbf{\omega}\| = 1\}$ be the parameter space for $\mathbf{\omega}$, and let $\mathbb{R}$ be the parameter space for $c$. Under some regularity conditions, the function 
    $$F(\mathbf{\omega},c)=(1-\pi)\mathbb{P}(\mathbf{\omega}^T\mathbf{Y}\leq c)-\pi \mathbb{P}(\mathbf{\omega}^T\mathbf{X}\leq c)$$
    attains at least one global maximum over $\Omega\times\mathbb{R}$.
\end{theorem}

To ensure the uniqueness of this maximizer, we introduce the following condition:
\begin{theorem}\label{theorem1}[Uniqueness]
    Assume $(\mathbf{\omega}, c) \in \Omega \times \mathbb{R}$ as defined above. The maximizer $(\mathbf{\omega}_0, c_0)$ is unique if, for any distinct pairs $(\mathbf{\omega}_1, c_1) \neq (\mathbf{\omega}_2, c_2)$, at least one of the following holds:
    \begin{equation*}
        \pi f_{\mathbf{\omega}_1^{T}\mathbf{X}}(c_1)\cdot \mathbb{E}[\mathbf{\omega}_2^{T}\mathbf{X}-c_2\mid\mathbf{\omega}_1^{T}\mathbf{X}=c_1]\neq (1-\pi)f_{\mathbf{\omega}_1^{T}\mathbf{Y}}(c_1)\cdot\mathbb{E}[\mathbf{\omega}_2^{T}\mathbf{Y}-c_2\mid \mathbf{\omega}_1^{T}\mathbf{Y}=c_1],
    \end{equation*}
    \begin{equation*}
        \pi f_{\mathbf{\omega}_2^{T}\mathbf{X}}(c_2)\cdot \mathbb{E}[\mathbf{\omega}_1^{T}\mathbf{X}-c_1\mid\mathbf{\omega}_2^{T}\mathbf{X}=c_2]\neq (1-\pi)f_{\mathbf{\omega}_2^{T}\mathbf{Y}}(c_2)\cdot\mathbb{E}[\mathbf{\omega}_1^{T}\mathbf{Y}-c_1\mid \mathbf{\omega}_2^{T}\mathbf{Y}=c_2].
    \end{equation*}
\end{theorem}
This condition does not assume any specific distributional form, such as normality, for the biomarkers in either group. While the condition may appear technical, exact equality of both sides is highly unlikely in practical applications, thereby supporting the identifiability of the optimal solution in a broad range of settings.

The uniqueness of $(\mathbf{\omega}_0,c_0)$ lays the groundwork for establishing several asymptotic properties of the estimator $(\widehat{\mathbf{\omega}},\widehat{c})$. In particular, if $(\mathbf{\omega}_0,c_0)$ is unique, then $(\widehat{\mathbf{\omega}},\widehat{c})$ inherits desirable properties such as consistency and asymptotic normality, provided certain conditions hold.

\begin{theorem}\label{theorem2}[Consistency]
    Suppose $\left(\mathbf{\omega}_0,c_0\right)$ is unique. Under some regularity conditions, the estimator $\left(\widehat{\mathbf{\omega}},\widehat{c}\right)$ is consistent, that is $\left(\widehat{\mathbf{\omega}},\widehat{c}\right) \xrightarrow{P} \left(\mathbf{\omega}_0,c_0\right)$ as $n_1,n_0 \to \infty$.
\end{theorem}
This result ensures that the estimated parameters converge to their true values as the sample sizes in both the diseased and healthy groups grow, thereby reinforcing the reliability of the proposed estimation procedure.

\begin{theorem}\label{theorem3}[Asymptotic Normality]
    Under standard regularity conditions, the estimator $\left(\widehat{\mathbf{\omega}}, \widehat{c}\right)$ is asymptotically normal: 
    \(
    \sqrt{n}\left(\left(\widehat{\mathbf{\omega}}, \widehat{c}\right) - \left(\mathbf{\omega}_0, c_0\right)\right) \xrightarrow{d} \mathcal{N}(0, \Sigma),
    \)
    where $\Sigma$ depends on the smoothness parameter $h$, the sample proportions $\rho_0 = \lim_{n \to \infty} n_0 / n$ and $\rho_1 = \lim_{n \to \infty} n_1 / n$, as well as the standard normal density $\phi(\cdot)$. Full details of $\Sigma$ are provided in supplementary material.
\end{theorem}

Asymptotic normality permits the construction of confidence intervals and hypothesis tests for the estimated parameters, offering a theoretical basis for their interpretation in practical applications. The corresponding regularity conditions and the proofs of Theorem \ref{theorem0}, \ref{theorem1}, \ref{theorem2} and \ref{theorem3} are provided in the supplementary material.

\section{The penalized weighted smoothed Youden index estimation}

In the previous section, we explore the properties of the smoothed weighted Youden index estimator without penalization. While informative, this approach may struggle in sparse scenarios where only a few biomarkers influence the test outcome, leading to overfitting, reduced model stability, and difficulty identifying relevant biomarkers. To address these issues, penalization techniques have been widely adopted to promote sparsity by shrinking irrelevant coefficients to zero, improving both interpretability and robustness. Notable methods include bridge regression \citep{frank1993statistical}, LASSO \citep{tibshirani1996regression}, and the SCAD penalty \citep{fan2001variable}. SCAD, in particular, offers unbiased estimation for large coefficients and sparse solutions, making it well-suited for biomarker selection \citep{lin2011selection}. 

Building on the previous framework, we incorporate the SCAD penalty into the smoothed weighted Youden index estimator. This penalized formulation combines the strengths of the weighted Youden index with SCAD’s sparsity-inducing properties, ensuring effective biomarker selection while preserving the desirable statistical properties of the estimator.

The SCAD penalty is defined by
\begin{equation*}
    p_{\lambda}^{\prime}(\omega)=\lambda\left\{\mathbb{I}(\omega\leq\lambda)+\frac{(a\lambda-\omega)_{+}}{(a-1)\lambda}\mathbb{I}(\omega>\lambda)\right\} \text{for}\;\text{some}\;a>2\;\text{and}\;\omega>0,
\end{equation*}
with $p_{\lambda}^{\prime}(0)=0$. Based on this penalty function, the proposed penalized smoothed weighted Youden index estimator is obtained by maximizing
\begin{equation}\label{SPWY}
    L_n(\mathbf{\omega},c)=(1-\pi)\frac{1}{n_0}\sum_{j=1}^{n_0}\Phi\left(\frac{c-\mathbf{\omega}^T\mathbf{Y}_j}{h}\right)-\pi\frac{1}{n_1}\sum_{i=1}^{n_1}\Phi\left(\frac{c-\mathbf{\omega}^{T}\mathbf{X}_i}{h}\right)-\sum_{t=1}^p p_{\lambda_n}(|\omega_t|),
\end{equation}
where $p$ is the dimension of $\mathbf{\omega}$. To satisfy the normalization constraint $\left\|\mathbf{\omega}\right\|=1$, we solve
\begin{equation} \label{eqn:optimization-max}(\widehat{\mathbf{\omega}}^{\text{SCAD}},\widehat{c}^{\text{SCAD}})=\arg\max_{\mathbf{\omega}\in\Omega,c\in\mathbb{R}}L_n(\mathbf{\omega},c),
\end{equation}
where $\Omega=\{\omega\in\mathbb{R}^P:\left
\|\omega\right\|=1\}$.

\subsection{Sampling properties and oracle properties}
In this subsection, we establish the asymptotic properties of the penalized smoothed weighted Youden index estimator. Our primary goals are to demonstrate its convergence rate, sparsity, and asymptotic normality under the SCAD penalty. These properties ensure that the estimator effectively identifies relevant biomarkers and achieves robust statistical performance.

Let $\mathbf{\omega}_0 = (\mathbf{\omega}_0^{(1)T}, \mathbf{\omega}_0^{(2)T})^T$, where $\mathbf{\omega}_0^{(2)} = 0$ without loss of generality. The vector $\mathbf{\omega}_0^{(1)}$ contains the non-zero components, and $s$ denotes their number. We show that the penalized estimator exists and converges at the rate $O_p(n^{-1/2} + a_n)$, where $a_n = \max_t\{p_{\lambda_n}^\prime(|\omega_{t0}|) : \omega_{t0} \neq 0\}$. This rate reflects the combined effects of sample size and the regularization penalty.

All technical proofs and regularity conditions are provided in the supplementary material. These proofs build on the framework of \citet{fan2001variable} with two key modifications: (i) the additional bias introduced by using a smooth approximation instead of the empirical distribution, and (ii) the normalization constraint $\|\mathbf{\omega}\| = 1$. Specifically, the discrepancy between the smoothed functions (e.g., involving $\phi(\cdot)$) and their empirical counterparts introduces an extra layer of approximation. Additionally, the penalty and normalization constraints apply only to $\mathbf{\omega}$, distinguishing our method from previous approaches.

\begin{theorem}\label{theorem4}[Convergence and Root-n Consistency]
    Suppose $\max_t\left\{p_{\lambda_n}^{\prime\prime}(|\omega_{t0}|):\omega_{t0}\neq 0\right\}\rightarrow 0$. Then there exists a local maximizer $(\widehat{\mathbf{\omega}}^{\text{SCAD}},\widehat{c}^{\text{SCAD}})$ of $L_n(\mathbf{\omega},c)$, satisfying $\left\|\widehat{\mathbf{\omega}}^{\text{SCAD}}\right\|=1$, such that:
    \begin{equation*}
        \left\|(\widehat{\mathbf{\omega}}^{\text{SCAD}},\widehat{c}^{\text{SCAD}})-(\mathbf{\omega}_0,c_0)\right\|=O_p(n^{-1/2}+a_n).
    \end{equation*}
\end{theorem}
This result indicates that under the SCAD penalty, the smoothed penalized weighted Youden index estimator is root-$n$ consistent when $\lambda_n\to 0$.

In addition to consistency, we also establish the sparsity and asymptotic normality of the proposed estimator, both of which are critical for effective biomarker selection in sparse settings. Sparsity ensures the exclusion of irrelevant biomarkers, while asymptotic normality enables valid statistical inference for the selected components.

Let $\Sigma=\text{diag}\left\{p_{\lambda_n}^{\prime\prime}\left(|\omega_{10}|\right),\cdots,p_{\lambda_n}^{\prime\prime}\left(|\omega_{s0}|\right)\right\}$, $\mathbf{b}=\left(p_{\lambda_n}^{\prime}\left(|\omega_{10}|\right)sgn\left(\omega_{10}\right),\cdots,p_{\lambda_n}^{\prime}\left(|\omega_{s0}|\right)sgn\left(\omega_{s0}\right)\right)^T$.

\begin{theorem}\label{theorem5}[Sparsity and Asymptotic Normality]
    Assume $\mathop{\text{liminf}}\limits_{n\rightarrow\infty}\mathop{\text{liminf}}\limits_{\theta\rightarrow 0^{+}}p^{\prime}_{\lambda_n}(\theta)/\lambda_n>0$. If $\lambda_n\to 0$ and $\sqrt{n}\lambda_n\rightarrow \infty$ as $n\to\infty$, with probability tending to 1, the $\sqrt{n}-$consistent local maximizer, $\widehat{\mathbf{\omega}}^{\text{SCAD}}=\left(\begin{array}{c}
        \widehat{\mathbf{\omega}}^{\text{SCAD}}_{(1)} \\
        \widehat{\mathbf{\omega}}^{\text{SCAD}}_{(2)}
    \end{array}\right)$, satisfies:

    (a) Sparsity: $\widehat{\mathbf{\omega}}^{\text{SCAD}}_{(2)}=0$.

    (b) Asymptotic normality:
    \begin{equation*}
        \sqrt{n}\left\{\left(\Sigma-H(\mathbf{\omega}_0)\right)\left(\widehat{\mathbf{\omega}}^{\text{SCAD}}_{(1)}-\mathbf{\omega}_0^{(1)}\right)+\mathbf{b}\right\}\xrightarrow{d}\mathcal{N}\left(0,\frac{\pi^2}{\rho_1}\text{Cov}\left(\mathbf{U}^{(1)}\right)+\frac{(1-\pi)^2}{\rho_0}\text{Cov}\left(\mathbf{V}^{(1)}\right)\right),
    \end{equation*}
    where $H(\mathbf{\omega}_0)=\frac{\partial^2 S_n\left(\mathbf{\omega}_0,c_0\right)}{\partial \mathbf{\omega}^{(1)T}\partial \mathbf{\omega}^{(1)}}$, $\mathbf{U}^{(1)}=\frac{1}{h}\phi\left(\frac{c-\mathbf{\omega}^T_0\mathbf{X}}{h}\right)\mathbf{X}^{(1)}$, $\mathbf{V}^{(1)}=-\frac{1}{h}\phi\left(\frac{c-\mathbf{\omega}^T_0\mathbf{Y}}{h}\right)\mathbf{Y}^{(1)}$.
\end{theorem}

These results confirm that the proposed penalized estimator achieves both accurate variable selection (by eliminating irrelevant components) and the asymptotic normality necessary for valid statistical inference in sparse biomarker analyses.

\section{Nonmonotone accelerate proximal gradient algorithm}
As established earlier, the proposed estimator maximizes a smoothed weighted Youden index with a nonconvex SCAD penalty. For algorithmic development, we reformulate this as an equivalent minimization problem:
\begin{equation} \label{model:opt}
 \min_{\mathbf{\omega},c}  F(\mathbf{v}) :=  f(\mathbf{\omega},c) + g(\mathbf{\omega},c),
\end{equation}
where $\mathbf{v} = [\mathbf{\omega}, c]$. The data-fitting term is $f(\mathbf{\omega},c)=\pi\frac{1}{n_1}\sum_{i=1}^{n_1}\Phi\left(\frac{c-\mathbf{\omega}^{T}\mathbf{X}_i}{h}\right) - (1-\pi)\frac{1}{n_0}\sum_{j=1}^{n_0}\Phi\left(\frac{c-\mathbf{\omega}^T\mathbf{Y}_j}{h}\right)$, and the regularization term is $g(\mathbf{\omega},c) = \lambda_1 p(\mathbf{\omega}) + \lambda_2 c^2$, where $p(\cdot)$ denotes the SCAD penalty and $\lambda_1,\lambda_2$ are regularization parameters. Note that the original formulation (\ref{SPWY}) includes no penalty on the cutoff parameter $c$. The term $\lambda_2c^2$ is introduced purely for numerical stability and to support convergence analysis. Since $\lambda_2$ is chosen to be small (e.g., $10^{-6}$), its effect on the final estimates is negligible and does not affect the statistical properties discussed earlier.

The objective function $F$ is both nonconvex and nonsmooth, due to the SCAD penalty and the nonconvexity of the data-fitting term. While accelerated proximal gradient (APG) algorithms are effective for smooth convex problems, their performance deteriorates when applied directly to nonsmooth and nonconvex objectives. Standard line search techniques, which rely on smoothness, often yield overly conservative step sizes or slow convergence in such settings. These challenges are further amplified in high-dimensional problems, where full gradient evaluations are computationally expensive. 

To address these challenges, we propose a nonmonotone accelerated proximal gradient (APG) algorithm that integrates proximal updates, adaptive step size selection, and nonmonotone descent control. By allowing controlled increases in the objective function and dynamically adjusting step sizes, the method improves per-iteration efficiency and enhances convergence stability in nonsmooth and nonconvex settings.

\subsection{Nonmonotone APG with line search}
We now detail the proposed nonmonotone APG algorithm. It integrates proximal gradient updates for nonsmooth regularization, Nesterov-style extrapolation for momentum-based acceleration, and a polynomial interpolation-based line search for adaptive step size selection. These elements are extended to the nonsmooth and nonconvex setting, allowing the algorithm to robustly handle composite objectives with complex structure.

The algorithm is initialized with $\mathbf{v}_1 = \mathbf{v}_0 = \mathbf{w}_0 = \mathbf{u}_1$, where $\mathbf{v}_1$ serves as the initial solution estimate. The extrapolated point $\mathbf{w}_0$ is computed via Nesterov-style momentum, while $\mathbf{u}_1$ is obtained through the proximal mapping, which generalizes gradient updates for nonsmooth functions. The extrapolation parameters are set as $t_1 = 1$ and $t_0 = 0$ and are updated in each iteration. The initial objective value is recorded as $c_1 = F(\mathbf{v}_1)$. To control nonmonotonicity, the algorithm uses two parameters: $\eta \in [0, 1]$, which allows controlled increases in the objective value, and $q_1 = 1$, a dynamic weight that tracks the moving average of recent function values to ensure long-term descent. Additionally, a descent threshold $\delta > 0$ is used to determine whether a candidate update satisfies the sufficient decrease condition, with larger values enforcing stricter descent.

At each iteration, the algorithm first computes the extrapolated point $\mathbf{w}_k$ based on the current and previous iterates. It then applies the Barzilai–Borwein (BB) method to determine an initial step size $\alpha_y$, alternating between two BB formulas to better adapt to local curvature \citep{dai2005projected}. This step size is refined via a polynomial interpolation-based backtracking line search, which ensures that the nonmonotone descent condition is met while computing the updated proximal point $\mathbf{u}_{k+1}$. If $F(\mathbf{u}_{k+1})$ meets the acceptance criterion, we set $\mathbf{v}_{k+1}=\mathbf{u}_{k+1}$. Otherwise, an alternative point $\mathbf{z}_{k+1}$ is computed using a capped BB step size $\alpha_x$ and another backtracking step. A selection rule is then applied to choose between $\mathbf{u}_{k+1}$ and $\mathbf{z}_{k+1}$, ensuring the better approximation is retained. Finally, the Nesterov parameter $t_k$ and the auxiliary variables $c_k$ and $q_k$ are updated to maintain momentum and manage nonmonotonicity. characteristics. These design choices jointly enable the algorithm to maintain stable and adaptive convergence throughout the optimization process. Full details are provided in Algorithms \ref{alg:sto-proxg} and \ref{alg-stepsize}.

\begin{algorithm} 
\caption{Nonmonotone APG with polynomial interpolation based backtracking line search}
\label{alg:sto-proxg}
\begin{algorithmic}[1] 
\State Initialize $\mathbf{v}_1 = \mathbf{v}_0 = \mathbf{w}_0 = \mathbf{u}_1, t_1 = 1, t_0 = 0, \eta \in [0, 1], \delta > 0, c_1 = F(\mathbf{v}_1), q_1 = 1.$
\For{$k = 1,2,3,\dots$}
    \State $\mathbf{w}_k = \mathbf{v}_k + \frac{t_{k-1}}{t_k} (\mathbf{u}_k - \mathbf{v}_k) + \frac{t_{k-1} - 1}{t_k}(\mathbf{v}_k - \mathbf{v}_{k-1})$,
    \State $\mathbf{s}_k = \mathbf{w}_k - \mathbf{w}_{k-1}, r_k = \nabla f(\mathbf{w}_k) - \nabla f(\mathbf{w}_{k-1})$,
\vspace{0.2cm}
    \State $\alpha_y = \begin{cases}
\frac{ |\mathbf{s}_k^T \mathbf{s}_k|}{|\mathbf{s}_k^T r_k|} & \text{if} \mod(k,2) = 1, \\
   \frac{|\mathbf{s}_k^T r_k|}{|r_k^T r_k|} & \mbox{else.}
    \end{cases}$
\vspace{0.2cm}
\State \label{linesearch1} Use Algorithm \ref{alg-stepsize} with intial stepsize $\alpha^{(0)} = \alpha_y$ to obtain stepsize $\alpha_{y*}$ and $\mathbf{u}_{k+1} = \text{prox}_{\alpha_{y*} g}(\mathbf{w}_k - \alpha_{y*} \nabla f(\mathbf{w}_k))$ such that $F(\mathbf{u}_{k+1}) \leq F(\mathbf{w}_k) - \delta \|\mathbf{u}_{k+1} - \mathbf{w}_k\|^2$ or $F(\mathbf{u}_{k+1}) \leq c_k - \delta \|\mathbf{u}_{k+1} - \mathbf{w}_k\|^2$.
    \If{$F(\mathbf{u}_{k+1}) \leq c_k - \delta \|\mathbf{u}_{k+1} - \mathbf{w}_k\|^2$}
        \State   \label{excute-type1}
            $\mathbf{v}_{k+1} = \mathbf{u}_{k+1} $,
    \Else
        \State  $\mathbf{s}_k = \mathbf{v}_k - \mathbf{w}_{k-1}, r_k = \nabla f(\mathbf{v}_k) - \nabla f(\mathbf{w}_{k-1})$, 
\vspace{0.2cm}
    \State $\alpha_x = \begin{cases}
\frac{|\mathbf{s}_k^T \mathbf{s}_k|}{|\mathbf{s}_k^T r_k|} & \text{if} \mod(k,2) = 1, \\
   \frac{|\mathbf{s}_k^T r_k|}{|r_k^T r_k|} & \mbox{else.}
    \end{cases}$
\vspace{0.2cm}
\State \label{linesearch2} Use Algorithm \ref{alg-stepsize}  with intial stepsize $\alpha^{(0)} = \alpha_{x}$ to obtain stepsize $\alpha_{x*}$ and $\mathbf{z}_{k+1} = \text{prox}_{\alpha_{x*} g}(\mathbf{v}_k - \alpha_{x*} \nabla f(\mathbf{v}_k))$ such that $F(\mathbf{z}_{k+1}) \leq c_k - \delta \|\mathbf{z}_{k+1} - \mathbf{v}_k\|^2$.

        \State \label{excute-type2} $\mathbf{v}_{k+1} = \begin{cases}
            \mathbf{u}_{k+1}, & \text{if } F(\mathbf{u}_{k+1}) \leq F(\mathbf{z}_{k+1}), \\
            \mathbf{z}_{k+1}, & \text{otherwise}.
        \end{cases}$
    \EndIf
    \State \label{t-update} $t_{k+1} = \frac{\sqrt{4(t_k)^2 + 1} + 1}{2}$,
    \State $q_{k+1} = \eta q_k + 1$,
    \State $c_{k+1} = \frac{\eta q_k c_k + F(\mathbf{v}_{k+1})}{q_{k+1}}$.
\EndFor
\end{algorithmic}
\end{algorithm}

\begin{algorithm} 
\caption{Polynomial interpolation-based backtracking algorithm}
\label{alg-stepsize}
\begin{algorithmic}[1]
\State \textbf{Notation:} Define $G_{\tilde{t}}(\mathbf{v}) :=  \frac{1}{\tilde{t}}\left(\mathbf{v} - \mathrm{prox}_{\tilde{t}g}(\mathbf{v} - \tilde{t} \nabla f(\mathbf{v}))\right)$ as the gradient mapping, and let $h(\alpha) := F(\mathbf{v} - \alpha G_{\tilde{t}}(\mathbf{v}))$. We use \( \|G_{\tilde{t}}(\mathbf{v})\|^2 \) as an approximation of \( h'(0) \).
\State \textbf{Input:} Initial step size $\alpha^{(0)}$, $0 < \tau_1 \leq \tau_2 < 1$, $h(0)$;
\State \textbf{Output:} Step size $\alpha_*$ satisfying the sufficient decrease condition
\State Set $i \gets 0$;
\Loop
    \If{$h(\alpha^{(i)}) \leq h(0) - c_1  \alpha^{(i)}\|G_{\alpha^{(i)} }(\mathbf{v}) \|^2  $ }
        \State $\alpha_* \gets \alpha^{(i)}$ and return,
    \EndIf
    \If{$h(\alpha^{(i)}) > h(0) - c_1  \alpha^{(i)}\|G_{\alpha^{(i)} }(\mathbf{v}) \|^2$ and $i = 0$}
        \State Compute $\tilde{\alpha}$ by \eqref{eqn:poly1} with $\alpha_1 = \alpha^{(i)}$,
    \EndIf
    \If{$h(\alpha^{(i)}) > h(0) - c_1  \alpha^{(i)}\|G_{\alpha^{(i)} }(\mathbf{v}) \|^2$ and $i > 0$}
        \State Compute $\tilde{\alpha}$ by \eqref{eqn:poly2} with $\alpha_1 = \alpha^{(i)}$ and $\alpha_2 = \alpha^{(i-1)}$,
    \EndIf
    \State $\alpha^{(i+1)} = \min(\max(\tilde{\alpha}, \tau_1 \alpha^{(i)}), \tau_2 \alpha^{(i)})$, \label{update-alpha}
    \State $i \gets i + 1$.
\EndLoop
\end{algorithmic}
\end{algorithm}

A central component of this algorithm is the interpolation-based step size selection. While effective in smooth optimization, its direct application to nonsmooth settings is hindered by the lack of well-defined directional derivatives. To address this, we introduce a gradient mapping that approximates descent directions by combining the gradient of the smooth component $f$ with the proximal operator of the nonsmooth penalty $g$. For a fixed parameter $\tilde{t}>0$, the gradient mapping is defined as
\(G_{\tilde{t}}(\mathbf{v}) :=  \frac{1}{\tilde{t}}\left(\mathbf{v} - \mathrm{prox}_{\tilde{t}g}(\mathbf{v} - \tilde{t} \nabla f(\mathbf{v}))\right),\)
which serves as a generalized descent direction. Using this direction, we define the auxiliary one-dimensional function: $h(\alpha):=F\left(\mathbf{v}-\alpha G_{\tilde{t}}(\mathbf{v})\right)$, and approximate the directional derivative $h^{\prime}(0)$ by $\|G_{\tilde{t}}(\mathbf{v})\|^2$. This allows step size selection through interpolation even when $F$ is nonsmooth. 

To compute an appropriate step size $\tilde{\alpha}$, we construct polynomial models of $h(\alpha)$ based on recent function evaluations. A quadratic model yields:
\begin{equation} \label{eqn:poly1}
        \tilde{\alpha} = \frac{-h'(0)\alpha_1^2}{2(h(\alpha_1) - h(0) - h'(0)\alpha_1)},
\end{equation}
where $h^{\prime}(0)$ is approximated using $\|G_{\tilde{t}}(\mathbf{v})\|^2$. A more accurate cubic model incorporating an additional point $h(\alpha_2)$ gives:
\begin{equation} \label{eqn:poly2}
    \tilde{\alpha} = \frac{-b + \sqrt{b^2 - 3ah'(0)}}{3a},
\end{equation}
with coefficients $a$ and $b$ derived from  
        \[
        \begin{bmatrix}
            a \\
            b 
        \end{bmatrix} 
        = \frac{1}{\alpha_1 - \alpha_2}
        \begin{bmatrix}
            \frac{1}{\alpha_1^2} & -\frac{1}{\alpha_2^2} \\
            -\frac{\alpha_2}{\alpha_1^2} & \frac{\alpha_1}{\alpha_2^2}
        \end{bmatrix}
        \begin{bmatrix}
            h(\alpha_1) - h(0) - h'(0)\alpha_1 \\
            h(\alpha_2) - h(0) - h'(0)\alpha_2 
        \end{bmatrix}.
        \]
This interpolation process can be iterated using updated evaluations of $h$ and $G_{\tilde{t}}$. To ensure numerical stability, we apply a safeguard $\alpha^{(i+1)}=\min\left(\max(\tilde{\alpha}, \tau_1 \alpha_2 ), \tau_2 \alpha_2\right)$, where $0<\tau_1\leq\tau_2<1$ are predefined constants that bound the step size away from extreme values.

Beyond step size selection, the gradient mapping also serves as a proxy for the stationarity condition. In nonsmooth and nonconvex problems, standard stationarity conditions are often undefined or difficult to evaluate. However, in weakly convex settings, the squared norm $\|G_{\tilde{t}}(\mathbf{v})\|^2$ provides a meaningful convergence measure and has been widely adopted in recent analyses \citep{davis2019stochastic}.

In summary, the use of gradient mapping enables interpolation-based line search in nonsmooth and nonconvex settings. Unlike classical backtracking schemes that rely on fixed reduction rules, our approach adaptively selects step sizes based on function values and gradient mapping information. This reduces the number of trial evaluations, improves step size quality, and better adapts to local curvature. Combined with proximal updates and nonmonotone control, the resulting algorithm achieves robust and efficient convergence. A theoretical analysis of its convergence properties is presented in the next section.

\subsection{Convergence analysis}

To analyze the convergence of the proposed algorithm, we consider a broader class of nonsmooth and nonconvex optimization problems beyond the specific model in \eqref{model:opt}. We assume that the smooth component $f(\mathbf{v})$ is proper and has $L$-Lipschitz continuous gradients, while the nonsmooth component $g(\mathbf{v})$ is proper, lower semicontinuous, and possibly nonconvex. The composite objective $F(\mathbf{v}) = f(\mathbf{v}) + g(\mathbf{v})$ is further assumed to be coercive, meaning $F(\mathbf{v})\to\infty$ as $\|\mathbf{v}\|\to\infty$. To handle the nonsmoothness of $g(\mathbf{v})$, we adopt subdifferential calculus and assume that both $f$ and $g$ admit well-defined Fréchet or limiting subdifferentials. The optimality condition of the proximal operator naturally involves these subdifferentials and plays a key role in our analysis. These assumptions ensure that the problem is well-posed and provide the foundation for the theoretical analysis. Definitions of the relevant concepts are provided in the supplementary material for reference.

Under these assumptions, the proposed algorithm satisfies a key descent property. The following lemma establishes that Algorithm \ref{alg:sto-proxg} ensures sufficient decrease, which is essential for the subsequent convergence results.


\begin{lemma}\label{lemma1} 
    In Algorithm \ref{alg:sto-proxg}, we have
    \begin{equation} \label{eqn:exist1}
        F(\mathbf{v}_k) \leq c_k \leq A_k, \quad A_k = \frac{1}{k} \sum_{i=1}^k F(\mathbf{v}_i), 
    \end{equation}
    and there exists $\alpha_x$ such that $\mathbf{z}_{k+1} = \operatorname{prox}_{\alpha_x g}(\mathbf{v}_k - \alpha_x \nabla f(\mathbf{v}_k))$ satisfies
    \begin{equation} \label{eqn:exist3}
        F(\mathbf{z}_{k+1}) \leq c_k - \delta \|\mathbf{z}_{k+1} - \mathbf{v}_k\|^2, 
    \end{equation}
    where $\delta$ is a small positive constant. Consequently, according to Line \ref{update-alpha} in Algorithm \ref{alg-stepsize}, the line search criterion is well-defined. 
\end{lemma}

Building on Lemma \ref{lemma1}, the following theorem shows that the generated sequence is bounded and any accumulation point satisfies a first-order stationarity condition.


\begin{theorem}\label{theorem 6}
Let $\Omega_1 = \{k_1, k_2, \dots\}$ and \( \Omega_2 = \{m_1, m_2, \dots\} \) denote the index sets corresponding to iterations in which Line \ref{excute-type1} and Line \ref{excute-type2} of Algorithm \ref{alg:sto-proxg} are executed, respectively. Under the assumptions stated above, the sequences \( \{\mathbf{v}_k\}, \{\mathbf{z}_k\} \) and \( \{\mathbf{w}_k\} \) generated by Algorithm \ref{alg:sto-proxg} are bounded. Moreover, the following properties hold: 
\begin{enumerate}
    \item If either \( \Omega_1 \) or \( \Omega_2 \) is finite, then every accumulation point $\mathbf{v}_*$ of $\{\mathbf{v}_k\}$ satisfies the stationarity condition: \( 0 \in \partial F(\mathbf{v}_*) \).
    \item If both \( \Omega_1 \) and \( \Omega_2 \) are infinite, then all accumulation points of the subsequences $\{\mathbf{v}_{k_j+1}\}$, $\{\mathbf{w}_{k_j}\}$, $\{\mathbf{v}_{m_j}\}$, $\{\mathbf{z}_{m_j+1}\}$, with $k_j\in\Omega_1$ and $m_j\in\Omega_2$, satisfy the stationarity condition: $0 \in \partial F(\cdot)$.
\end{enumerate}
\end{theorem}

The proofs of Lemma \ref{lemma1} and Theorem \ref{theorem 6} are provided in the supplementary material.

\section{Simulation}

This section presents two simulation studies. The first evaluates the finite-sample performance of the proposed method in comparison with lasso-logistic regression, a widely recognized benchmark for biomarker selection and combination. Other methods were excluded due to limitations such as computational inefficiency or restrictive assumptions. The second study examines the convergence speed of the proposed algorithm, comparing it with the baseline accelerated proximal gradient (APG) algorithm and its variant, MAPG \citep{li2015accelerated}, as employed by \citet{salaroli2023pye}. The APG algorithm serves as a foundational benchmark, while MAPG offers a practical comparison for assessing efficiency.

In both simulation studies, the penalization parameter $\lambda$ is selected from the set $(10,5,1,$ $0.5,0.1,0.05,0.01,0.005)$ via cross-validation on a separate simulated dataset, optimizing the weighted Youden index on the validation set. Following \citet{fan2001variable}, the shape parameter of the SCAD penalty function is set to 3.7. The bandwidth $h$ is determined as $(n_1n_0)^{-0.1}$, in accordance with the recommendations of \citet{vexler2006note}.

\subsection{Performance evaluation of the proposed method} \label{sec5-1}

This subsection compares the performance of the proposed method with that of the lasso-logistic method. In the lasso-logistic approach, the linear coefficients are first estimated, and the cutoff point is subsequently chosen to maximize the weighted Youden index. We evaluate performance under three simulation scenarios: (1) data generated from a logistic regression model; (2) data generated from a highly complex single-index model; and (3) data generated from a highly complex single-index model with the number of biomarkers exceeding the sample size. For each scenario, two values of the weight parameter $\pi$ in the weighted Youden index are considered: $\pi=0.5$, corresponding to the standard Youden index, and $\pi=0.6$, which assigns greater weight to sensitivity. In each setting, 1,000 datasets are generated.

In the first scenario, data are generated from a logistic regression model with the true coefficient vector $\mathbf{\omega}=(4,0,6,0,0,7,0,8,0,0)^{T}$, and each component of $\mathbf{T}$ is independently drawn from a standard normal distribution. The total sample sizes considered are 400, 1000, and 2000, with each dataset randomly split into training and testing sets of equal size. Table \ref{youden_setting1} summarizes the training and testing performance, detection rate, and shrinkage accuracy of the classification rules obtained by both methods for $\pi = 0.5$ and $\pi = 0.6$. Here, detection rate is defined as the proportion of truly nonzero coefficients correctly identified, while shrinkage accuracy refers to the proportion of truly zero coefficients correctly estimated as zero. Estimates of individual coefficients and cutoff values are provided in the supplementary material. Although the logistic model is correctly specified in this setting, the proposed method achieves comparable or slightly superior performance to lasso-logistic in terms of the weighted Youden index. Moreover, our method demonstrates improved variable selection by more effectively shrinking irrelevant coefficients toward zero.


\begin{table}
\caption{Training and testing performance, detection rate, and shrinkage accuracy under the first data-generation scenario, evaluated using the weighted Youden index with $\pi=0.5$ (standard Youden index) and $\pi=0.6$. The detection rate is defined as the proportion of true non-zero parameters correctly identified, while shrinkage accuracy refers to the proportion of true zero parameters correctly estimated as zero.  \label{youden_setting1}}
\begin{center}
\begin{tabular}{lllllllll}
    \Hline
     & \multicolumn{4}{@{}l}{Mean of weighted Youden index} & \multicolumn{2}{@{}l}{Detection rate} & \multicolumn{2}{@{}l}{Shrinkage accuracy} \\ 
    & \multicolumn{2}{@{}l}{Training set} & \multicolumn{2}{@{}l}{Testing set} \\
    Sample size & Ours & Logit & Ours & Logit & Ours & Logit & Ours & Logit  \\
    \hline
    \multicolumn{9}{@{}c}{$\pi=0.5$} \\
    400 & 0.9720 & 0.9796 & 0.9252 & 0.9186 & 1 & 1 & 41.10\% & 27.70\% \\
         
    1000 & 0.9637 & 0.9667 & 0.9441 & 0.9390 & 1 & 1 & 61.14\% & 40.98\% \\
         
    2000 & 0.9613 & 0.9629 & 0.9508 & 0.9475 & 1 & 1 & 77.33\% & 60.08\% \\

    \multicolumn{9}{@{}c}{$\pi=0.6$} \\
    400 & 0.9659 & 0.9612 & 0.9246 & 0.9218 & 1 & 1 & 54.60\% & 52.72\% \\
         
    1000 & 0.9604 & 0.9629 & 0.9437 & 0.9420 & 1 & 1 & 76.58\% & 41.78\% \\
         
    2000 & 0.9568 & 0.9609 & 0.9494 & 0.9460 & 1 & 1 & 88.86\% & 56.30\% \\
    \hline
\end{tabular}
\end{center}
\end{table}

In the second scenario, the true coefficient vector is set to $\mathbf{\omega}=(-5, -4, -4.5, 3.5, 0,0,0,0,0,0)^T$. The first four biomarkers are generated from chi-square, gamma, exponential, and $t$ distributions, respectively, with dependencies introduced via a Gaussian copula. The remaining six biomarkers are independently drawn from standard normal distributions. To generate the disease status, a complex and asymmetric link function is used, with additional random noise added to increase the complexity of the data-generating process. The total sample sizes considered are 400, 1000, and 2000, with each dataset randomly split into training and testing sets of equal size. Results for $\pi = 0.5$ and $\pi = 0.6$ are reported in Table \ref{youden_setting2}.

The results show that the proposed method substantially outperforms lasso-logistic in terms of the weighted Youden index across both values of $\pi$. Regarding detection rate and shrinkage accuracy, for $\pi = 0.5$, our method exhibits slightly lower detection rates in some cases but substantially higher shrinkage accuracy compared to lasso-logistic. For $\pi = 0.6$, the proposed method performs slightly worse in small-sample settings (e.g., $n=400$), but as the sample size increases, it surpasses lasso-logistic on both metrics.

Interestingly, the relationship between sample size and performance is not always monotonic. Both detection rate and shrinkage accuracy exhibit fluctuations, which may stem from data randomness. These irregular patterns are consistent with phenomena reported in other penalized regression frameworks \citep{fan2001variable, zou2005regularization}. Notably, under $\pi=0.5$, as the sample size increases, lasso-logistic achieves a significantly higher detection rate. However, this improvement comes at the cost of decreased shrinkage accuracy, suggesting that lasso-logistic may be overfitting to noise by selecting irrelevant variables.

\begin{table}
\caption{Training and testing performance, detection rate, and shrinkage accuracy under the second data-generation scenario, evaluated using the weighted Youden index with $\pi=0.5$ (standard Youden index). The detection rate is defined as the proportion of true non-zero parameters correctly identified, while shrinkage accuracy refers to the proportion of true zero parameters correctly estimated as zero.  \label{youden_setting2}}
\begin{center}
\begin{tabular}{lllllllll}
    \Hline
     & \multicolumn{4}{@{}l}{Mean of weighted Youden index} & \multicolumn{2}{@{}l}{Detection rate} & \multicolumn{2}{@{}l}{Shrinkage accuracy} \\ 
    & \multicolumn{2}{@{}l}{Training set} & \multicolumn{2}{@{}l}{Testing set} \\
    Sample size & Ours & Logit & Ours & Logit & Ours & Logit & Ours & Logit  \\
    \hline
    \multicolumn{9}{@{}c}{$\pi=0.5$} \\
    400 & 0.4808 & 0.4416 & 0.4200 & 0.3342 & 80.20\% & 75.92\% & 59.03\% & 65.33\% \\
         
    1000 & 0.4914 & 0.4078 & 0.4644 & 0.3385 & 95.80\% & 86.12\% & 65.45\% & 46.72\% \\
         
    2000 & 0.4986 & 0.4079 & 0.4913 & 0.3740 & 88.05\% & 90.35\% & 74.60\% & 33.63\% \\
    \multicolumn{9}{@{}c}{$\pi=0.6$} \\
    400 & 0.5364 & 0.2043 & 0.4060 & 0.2062 & 78.45\% & 90.80\% & 25.86\% & 34.82\% \\
         
    1000 & 0.4908 & 0.2019 & 0.4436 & 0.2086 & 82.27\% & 81.60\% & 40.78\% & 37.13\% \\
         
    2000 & 0.4605 & 0.2010 & 0.4531 & 0.2115 & 91.10\% & 95.10\% & 48.91\% & 50.43\%  \\
    \hline
\end{tabular}
\end{center}
\end{table}

In the third simulation, data are generated under a high-dimensional setting with 500 biomarkers. The true coefficient vector assigns nonzero values only to the first 10 biomarkers, specifically $(-5,4.5,-4.5,3.5,-3,2.5,-2,1.5,-1,0.5)$, while the remaining 490 coefficients are set to zero. The biomarkers are generated from a multivariate Gaussian distribution with an AR(1) correlation structure. Additional structured perturbations are introduced to the latter 490 biomarkers to increase the difficulty of variable selection. The disease status is generated by applying a complex, asymmetric, and nonlinear link function to the linear predictor, with additional random noise incorporated to increase model complexity. We consider total sample sizes of 200, 400, and 600, ensuring that in all cases the training sample size is smaller than the number of biomarkers.

Results for $\pi=0.5$ and $\pi=0.6$ are reported in Table \ref{youden_setting3}. In this challenging high-dimensional setting with substantial noise and nonlinear perturbations, the proposed method consistently outperforms lasso-logistic in terms of the weighted Youden index across all settings. In addition, it achieves significantly better detection rates and shrinkage accuracy by more effectively eliminating irrelevant coefficients and more accurately identifying the true nonzero components.


\begin{table}
\caption{Training and testing performance, detection rate, and shrinkage accuracy under the third data-generation scenario, evaluated using the weighted Youden index with $\pi=0.5$ (standard Youden index). The detection rate is defined as the proportion of true non-zero parameters correctly identified, while shrinkage accuracy refers to the proportion of true zero parameters correctly estimated as zero.  \label{youden_setting3}}
\begin{center}
\begin{tabular}{lllllllll}
    \Hline
     & \multicolumn{4}{@{}l}{Mean of weighted Youden index} & \multicolumn{2}{@{}l}{Detection rate} & \multicolumn{2}{@{}l}{Shrinkage accuracy} \\ 
    & \multicolumn{2}{@{}l}{Training set} & \multicolumn{2}{@{}l}{Testing set} \\
    Sample size & Ours & Logit & Ours & Logit & Ours & Logit & Ours & Logit  \\
    \hline

    \multicolumn{9}{@{}c}{$\pi=0.5$} \\
         
    200 & 0.7004 & 1.0000 & 0.1209 & 0.0149 & 67.23\% & 45.77\% & 67.63\% & 65.13\%  \\
    
    400 & 0.5085 & 1.0000 & 0.1382 & 0.0322 & 75.76\% & 46.59\% & 71.96\% & 64.51\% \\
         
    600 & 0.4118 & 0.9963 & 0.1457 & 0.0497 & 78.65\% & 45.76\% & 75.17\% & 63.40\% \\

    \multicolumn{9}{@{}c}{$\pi=0.6$} \\

    200 & 0.2955 & 0.2126 & 0.2133 & 0.1991 & 86.70\% & 39.80\% & 89.08\% & 81.89\%  \\
    
    400 & 0.3626 & 0.2058 & 0.2220 & 0.1993 & 95.67\% & 47.11\% & 41.55\% & 64.36\% \\
         
    600 & 0.2204 & 0.2043 & 0.2095 & 0.1996 & 99.21\% & 46.50\% & 94.78\% & 62.91\% \\
    \hline
\end{tabular}
\end{center}
\end{table}


\subsection{Convergence analysis of the proposed algorithm}

\begin{figure}
\begin{center}
\begin{minipage}[b]{0.24\linewidth}
\centering
\includegraphics[width=\linewidth]{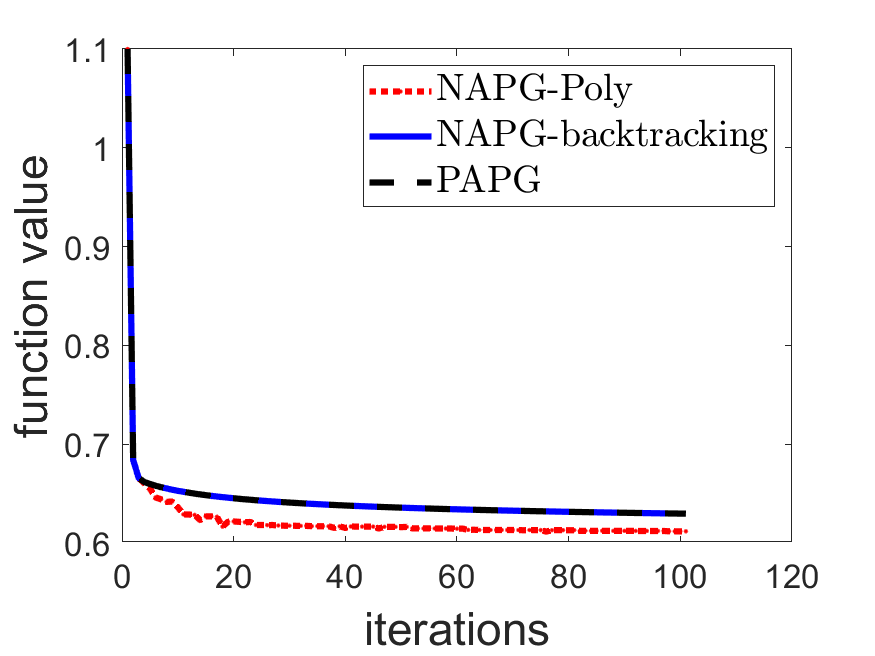}
\end{minipage}
\hfill
\begin{minipage}[b]{0.24\linewidth}
\centering
 \hfill\includegraphics[width=\linewidth]{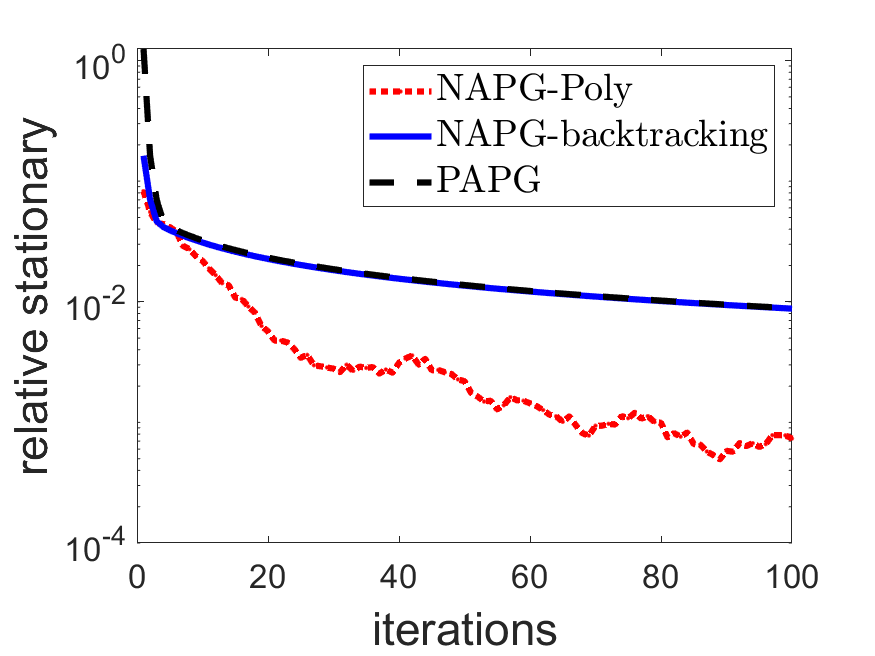}
\end{minipage}
\hfill
\begin{minipage}[b]{0.24\linewidth}
\centering
 \hfill\includegraphics[width=\linewidth]{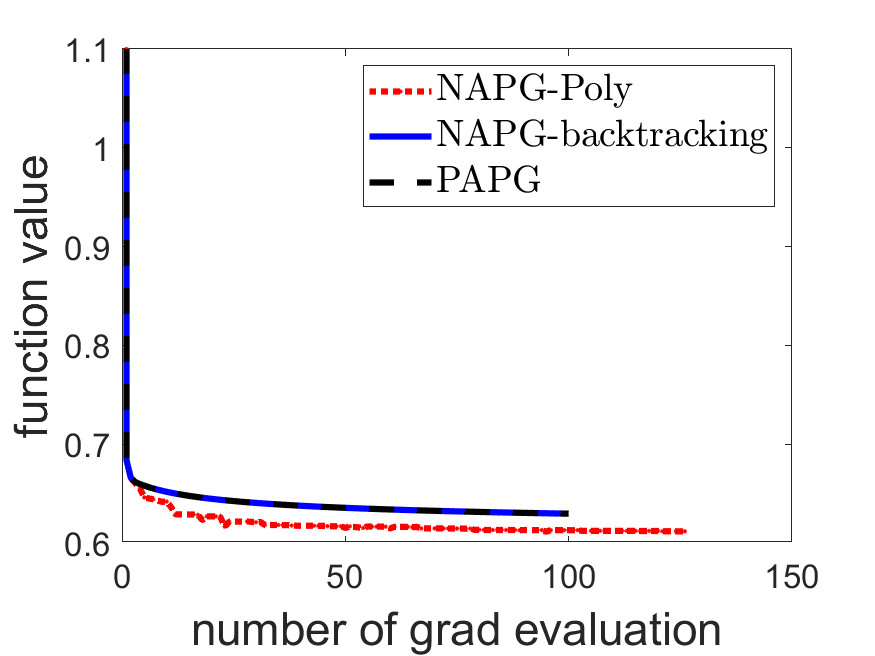}
\end{minipage}
\hfill
\begin{minipage}[b]{0.24\linewidth}
\centering
 \hfill\includegraphics[width=\linewidth]{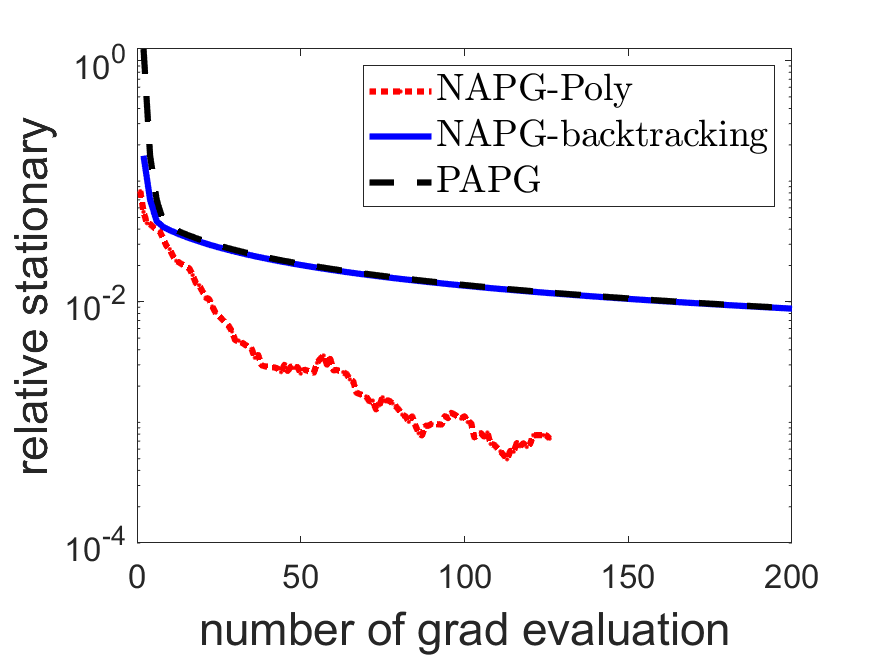}
\end{minipage}
\end{center}
\caption{Function values and relative stationary over iterations and gradient evaluations on simulated dataset. \label{fig:compare-simulate}}
\end{figure}

In this subsection, we compare the convergence speed of three algorithms: the proposed non-monotone accelerated proximal gradient algorithm with polynomial interpolation (NAPG-Poly, dotted line), the baseline accelerated proximal gradient (APG) method (PAPG, dashed line), and the APG algorithm with backtracking line search used in \citet{salaroli2023pye} (NAPG-backtracking, solid line). 
The comparison evaluates two key metrics: (1) function values, representing the objective values, and (2) relative stationarity, measuring the residual norm
$\| \mathbf{v}_k - \text{prox}_{\lambda_1 p}(\mathbf{v}_k -  \nabla f(\mathbf{v}_k) ) \|$.
These metrics assess convergence speed by capturing both the reduction in the objective function and proximity to stationarity. The horizontal axis in the plots represents iterations and the number of gradient evaluations, assessing both convergence speed and computational cost. Results, presented in Figure \ref{fig:compare-simulate}, highlight the computational efficiency and numerical stability of the three algorithms.

The first two panels in Figure \ref{fig:compare-simulate} depict how the function value and relative stationarity decrease with iterations. NAPG-Poly converges significantly faster than NAPG-backtracking and PAPG, achieving lower objective values and residual norms with fewer iterations. This demonstrates its effectiveness in reducing the objective function and improving stationarity at an accelerated pace. NAPG-backtracking exhibits smoother and more stable convergence than PAPG but is slower than NAPG-Poly, particularly in the early stages of optimization. PAPG converges at the slowest rate, reflecting the limitations of its fixed step size strategy in non-monotone and adaptive optimization contexts. The last two panels focus on convergence with respect to the number of gradient evaluations, which dominate the computational cost. NAPG-Poly demonstrates superior efficiency, requiring fewer gradient evaluations to achieve comparable reductions in function values and residual norms. Its polynomial interpolation strategy minimizes unnecessary gradient computations, making it particularly advantageous for large-scale optimization problems.

Overall, these results confirm the reliability and efficacy of the proposed NAPG-Poly, as described in Algorithm \ref{alg:sto-proxg}. Its rapid reduction in objective values, improved stationarity, and reduced gradient evaluations make it a robust and computationally efficient choice for tackling complex optimization problems.

\section{Real data example}

This section employs real-world data sourced from the national registry of the China Type II Inflammatory Skin Disease Clinical Research and Standardized Diagnosis and Treatment Project. Patients diagnosed with atopic dermatitis (AD) were identified according to the Williams Diagnostic Criteria in the dermatology departments of participating hospitals across 30 of China’s 34 provinces. A total of 205 hospitals contributed to the dataset, including two dedicated exclusively to pediatric healthcare \citep{zhao2024associations}.

This study includes 75 biomarkers, encompassing clinical features of skin lesions, lesion manifestations, symptoms, and signs. After excluding incomplete records, the dataset includes 34,957 samples, which are randomly split into equal-sized training and testing sets. The penalization parameter, $\lambda$, is selected from the set $(10, 5, 1, 0.5, 0.1, 0.05, 0.01, 0.005)$ via cross-validation on the training set, with the value that maximizes the weighted Youden index being chosen. The shape parameter $a_n$ for the SCAD penalty function is set to 3.7, and based on expert recommendations, the weight parameter $\pi$ is set to 0.6. The final selected penalization terms are $\lambda = 0.005$ for the proposed method and $\lambda = 1$ for the lasso-logistic method.

We assess the performance of both our proposed method and the lasso-logistic method on this real-world dataset, focusing on their weighted Youden index values and shrinkage rates. Additionally, we compare the convergence speeds of our non-monotone accelerated proximal gradient algorithm with polynomial interpolation (NAPG-Poly) against the APG algorithm with backtracking line search (NAPG-backtracking) and the baseline accelerated proximal gradient (PAPG) algorithm.

\begin{table}
\caption{Training and testing performance and zero shrinkage proportion in the real data example, with the objective function based on the maximization of the weighted Youden index ($\pi=0.6$). \label{youden_realdata}}
\begin{center}
\begin{tabular}{llll}
    \Hline
    & \multicolumn{2}{@{}l}{Weighted Youden index} & \\
    & Training set & Testing set & Zero shrinkage proportion \\\hline
         Ours & 0.7877 & 0.7819 & 58.67\% \\
         Logit & 0.7907 & 0.7852 & 4\% \\
    \hline
\end{tabular}
\end{center}
\end{table}

\begin{figure}
\centering
\begin{minipage}[b]{0.24\linewidth}
\centering
\includegraphics[width=\linewidth]{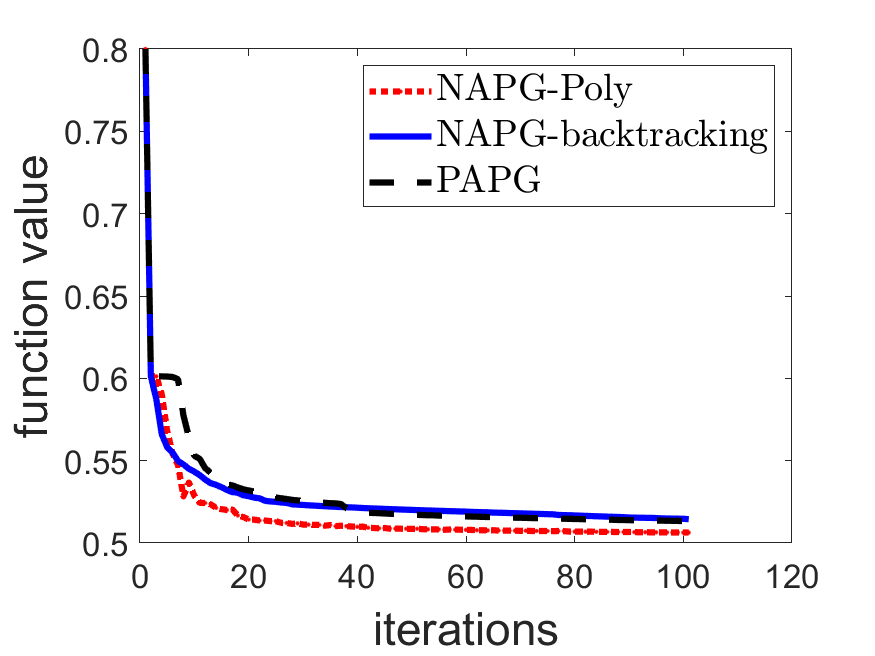}
\end{minipage}
\hfill
\begin{minipage}[b]{0.24\linewidth}
\centering
 \hfill\includegraphics[width=\linewidth]{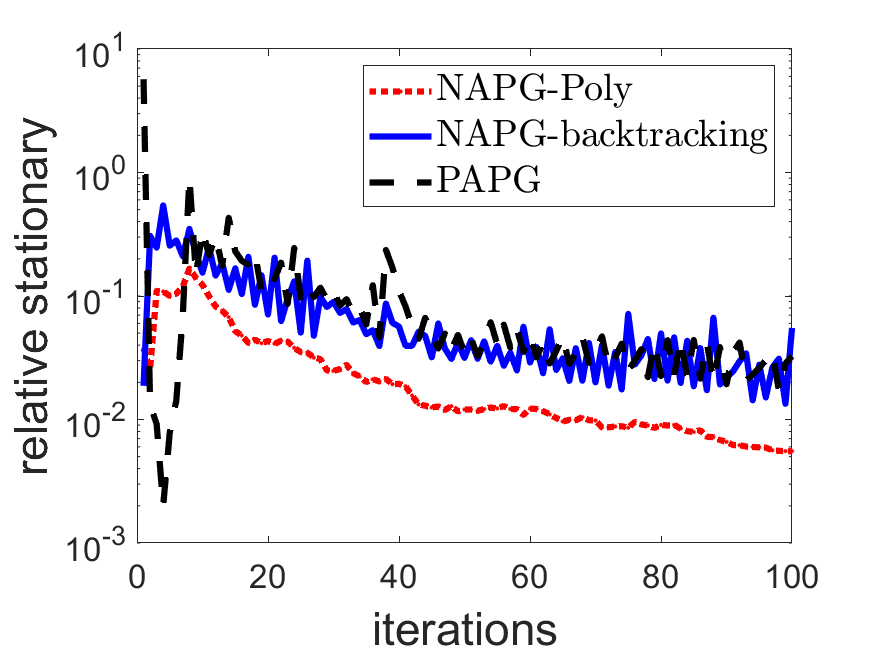}
\end{minipage}
\hfill
\begin{minipage}[b]{0.24\linewidth}
\centering
 \hfill\includegraphics[width=\linewidth]{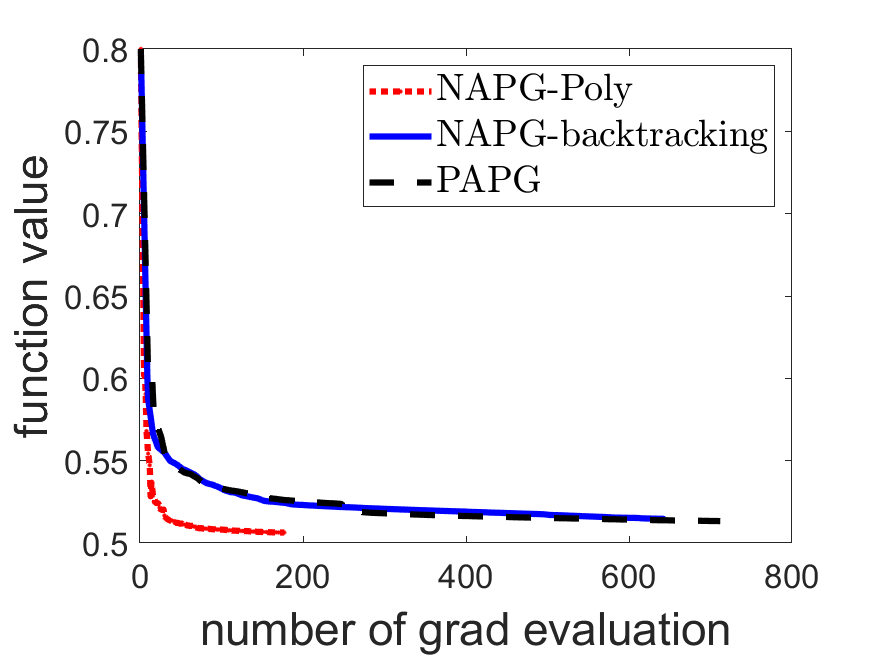}
\end{minipage}
\hfill
\begin{minipage}[b]{0.24\linewidth}
\centering
 \hfill\includegraphics[width=\linewidth]{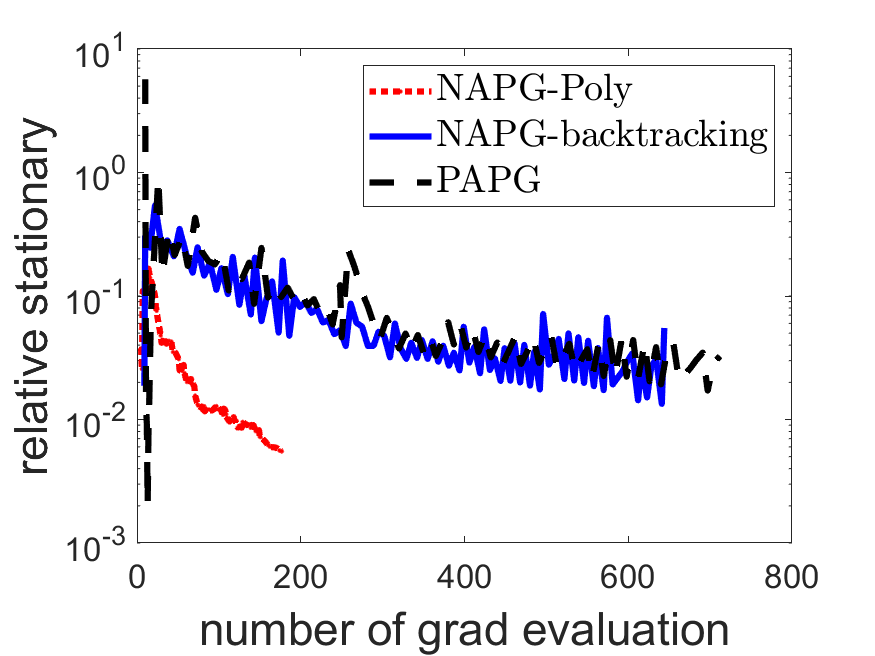}
\end{minipage}
\caption{Function values and relative stationary over iterations and gradient evaluations on real data example.}
\label{fig:compare-realdata}
\end{figure}

Table~\ref{youden_realdata} presents the performance results, and Figure~\ref{fig:compare-realdata} illustrates the convergence behavior. The weighted Youden index values attained by our proposed method are comparable to those of the lasso-logistic method, differing by only about 0.003 on the testing set. However, our method demonstrates a much higher rate of zero shrinkage, reducing 42 biomarker coefficients to zero compared to only 3 in the lasso-logistic approach, thus highlighting its superior sparsity and variable selection capabilities. In terms of convergence speed, our algorithm clearly outperforms NAPG-backtracking and PAPG by achieving lower objective values and residual norms using fewer iterations and gradient evaluations (Figure~\ref{fig:compare-realdata}), underscoring its computational efficiency in real-world settings where computational cost is critical. Overall, these findings show that our framework is not only competitive with the lasso-logistic method in terms of weighted Youden index performance but also significantly surpasses it in sparsity and computational efficiency, demonstrating its practical utility and robustness.

\section{Discussion}

In this article, we introduce a novel framework for selecting and combining multiple biomarkers to maximize the weighted Youden index, addressing key challenges in diagnostic biomarker analysis. Our approach features a smoothed weighted Youden index estimator and its penalized variant, leveraging the SCAD penalty for robust variable selection and sparsity. To overcome the nonconvex nature of the objective function and the additional nonsmoothness imposed by the penalty term, we propose a specialized algorithm optimized for computational efficiency. This algorithm is not restricted to our specific problem alone; it can also be applied to optimize a broad range of other nonconvex objectives. Through extensive simulations, our framework demonstrates superior performance compared to the LASSO-logistic model and showed efficiency gains when benchmarked against a baseline algorithm.

In practical analyses, a perfect gold standard may not always be available, and researchers may have to rely on an imperfect gold standard. Under the assumption that the imperfect gold standard and the biomarkers are conditionally independent given the true disease status, our framework still applies if the weight in the weighted Youden index is set to 0.5. In this situation, the pseudo Youden index derived from the imperfect gold standard remains a monotonic function of the true Youden index based on the actual disease status \citep{sun2024biomarker}, allowing direct estimation of the linear combination coefficients and the cutoff point by maximizing the pseudo Youden index or its penalized variant. However, this monotonic relationship no longer holds when the weight differs from 0.5. Developing methods for biomarker selection and combination under these circumstances—i.e., when a perfect gold standard is unavailable and the weight parameter deviates from 0.5—remains an important avenue for future research.



\begin{thebibliography}{}
\bibitem[\protect\citeauthoryear{Becker et al.}{2011}]{becker2011elastic} 
    Becker, N., Toedt, G., Lichter, P., and Benner, A. (2011). Elastic scad as a novel penalization method for svm classification tasks in high-dimensional data. BMC bioinformatics, 12:1–13.
\bibitem[\protect\citeauthoryear{Breiman}{1996}]{breiman1996heuristics} 
    Breiman, L. (1996). Heuristics of instability and stabilization in model selection. The annals of statistics, 24(6):2350–2383.
\bibitem[\protect\citeauthoryear{Dai and Fletcher}{2005}]{dai2005projected} 
    Dai, Y.-H. and Fletcher, R. (2005). Projected barzilai-borwein methods for large-scale box-constrained quadratic programming. Numerische Mathematik, 100(1):21–47.
\bibitem[\protect\citeauthoryear{Davis and Drusvyatskiy}{2019}]{davis2019stochastic} 
    Davis, D. and Drusvyatskiy, D. (2019). Stochastic model-based minimization of weakly convex functions. SIAM Journal on Optimization, 29(1):207–239.
\bibitem[\protect\citeauthoryear{Fan and Li}{2001}]{fan2001variable} 
    Fan, J. and Li, R. (2001). Variable selection via nonconcave penalized likelihood and its oracle properties. Journal of the American statistical Association, 96(456):1348–1360.
\bibitem[\protect\citeauthoryear{Frank and Friedma}{1993}]{frank1993statistical} 
    Frank, L. E. and Friedman, J. H. (1993). A statistical view of some chemometrics regression tools. Technometrics, 35(2):109–135.
\bibitem[\protect\citeauthoryear{Hsu et al.}{2014}]{hsu2014biomarker} 
    Hsu, M.-J., Chang, Y.-C. I., and Hsueh, H.-M. (2014). Biomarker selection for medical diagnosis using the partial area under the roc curve. BMC research notes, 7:1–15.
\bibitem[\protect\citeauthoryear{Li et al.}{2013}]{li2013weighted} 
    Li, D.-l., Shen, F., Yin, Y., Peng, J.-x., and Chen, P.-y. (2013). Weighted youden index and its two-independent-sample comparison based on weighted sensitivity and specificity. Chinese medical journal, 126(6):1150–1154.
\bibitem[\protect\citeauthoryear{Li and Lin}{2015}]{li2015accelerated} 
    Li, H. and Lin, Z. (2015). Accelerated proximal gradient methods for nonconvex programming. Advances in neural information processing systems, 28.
\bibitem[\protect\citeauthoryear{Li et al.}{2022}]{li2022applying} 
    Li, Y., Lu, F., and Yin, Y. (2022). Applying logistic lasso regression for the diagnosis of atypical crohn’s disease. Scientific Reports, 12(1):11340.
\bibitem[\protect\citeauthoryear{Lin et al.}{2011}]{lin2011selection} 
    Lin, H., Zhou, L., Peng, H., and Zhou, X.-H. (2011). Selection and combination of biomarkers using roc method for disease classification and prediction. Canadian Journal of Statistics, 39(2):324–343.
\bibitem[\protect\citeauthoryear{Ma and Huang}{2005}]{ma2005regularized} 
    Ma, S. and Huang, J. (2005). Regularized ROC method for disease classification and biomarker selection with microarray data. Bioinformatics, 21(24):4356–4362.
\bibitem[\protect\citeauthoryear{Ma and Huang}{2007}]{ma2007combining} 
    Ma, S. and Huang, J. (2007). Combining multiple markers for classification using ROC. Biometrics, 63(3):751–757.
\bibitem[\protect\citeauthoryear{Ma et al.}{2017}]{ma2017concordance} 
    Ma, Y., Li, Y., and Lin, H. (2017). Concordance measure-based feature screening and variable selection. Statistica Sinica, pages 1967–1985.
\bibitem[\protect\citeauthoryear{Neyman and Pearson}{1933}]{neyman1933ix} 
    Neyman, J. and Pearson, E. S. (1933). Ix. on the problem of the most efficient tests of statistical hypotheses. Philosophical Transactions of the Royal Society of London. Series A, Containing Papers of a Mathematical or Physical Character, 231(694-706):289–337.
\bibitem[\protect\citeauthoryear{Pepe et al.}{2006}]{pepe2006combining} 
    Pepe, M. S., Cai, T., and Longton, G. (2006). Combining predictors for classification using the area under the receiver operating characteristic curve. Biometrics, 62(1), 221–229.
\bibitem[\protect\citeauthoryear{Salaroli and del Carmen Pardo}{2023}]{salaroli2023pye} 
    Salaroli, C. J. and del Carmen Pardo, M. (2023). Pye: A penalized youden index estimator for selecting and combining biomarkers in high-dimensional data. Chemometrics and Intelligent Laboratory Systems, 236:104786.
\bibitem[\protect\citeauthoryear{Sun et al.}{2024}]{sun2024biomarker} 
    Sun, A., Li, Y., and Zhou, X.-H. (2024). Biomarker combination based on the youden index with and without gold standard. arXiv preprint arXiv:2412.17471.
\bibitem[\protect\citeauthoryear{Sun et al.}{2017}]{sun2017avc} 
    Sun, L., Wang, J., and Wei, J. (2017). Avc: Selecting discriminative features on basis of auc by maximizing variable complementarity. BMC bioinformatics, 18:73–89.
\bibitem[\protect\citeauthoryear{Tibshirani}{1996}]{tibshirani1996regression} 
    Tibshirani, R. (1996). Regression shrinkage and selection via the lasso. Journal of the Royal Statistical Society Series B: Statistical Methodology, 58(1):267–288.
\bibitem[\protect\citeauthoryear{Unger et al.}{2006}]{unger2006diagnostic} 
    Unger, N., Pitt, C., Schmidt, I. L., Walz, M. K., Schmid, K. W., Philipp, T., et al. (2006). Diagnostic value of various biochemical parameters for the diagnosis of pheochromocytoma in patients with adrenal mass. European Journal of Endocrinology, 154(3):409–417.
\bibitem[\protect\citeauthoryear{Vexler et al.}{2006}]{vexler2006note} 
    Vexler, A., Liu, A., Schisterman, E. F., and Wu, C. (2006). Note on distribution-free estimation of maximum linear separation of two multivariate distributions. Nonparametric Statistics, 18(2):145–158.
\bibitem[\protect\citeauthoryear{Yu and Park}{2014}]{yu2014aucpr} 
    Yu, W. and Park, T. (2014). Aucpr: an auc-based approach using penalized regression for disease prediction with high-dimensional omics data. BMC genomics, 15:1–12.
\bibitem[\protect\citeauthoryear{Zhao et al.}{2024}]{zhao2024associations} 
    Zhao, J., Zhang, Z., Chen, H., Dou, X., Zhao, Z., Liu, L., Wang, Y., and Li, H. (2024). Associations of demographics, aggravating factors, comorbidities, and treatments with atopic dermatitis severity in china: A national cross-sectional study. Chinese Medical Journal, pages 10–1097.
\bibitem[\protect\citeauthoryear{Zou and Hastie}{2005}]{zou2005regularization} 
    Zou, H. and Hastie, T. (2005). Regularization and variable selection via the elastic net. Journal of the Royal Statistical Society Series B: Statistical Methodology, 67(2):301–320.





\end{thebibliography}
\end{document}